\documentclass[conference]{IEEEtran}

\usepackage{multirow}
\usepackage{flushend}

\ifCLASSINFOpdf
  \usepackage[pdftex]{graphicx}
\else
  \usepackage[dvips]{graphicx}
\fi

\ifCLASSOPTIONcompsoc
  \usepackage[font=normalsize,labelfont=sf,textfont=sf]{subfig}
\else
  \usepackage[font=footnotesize]{subfig}
\fi

\begin{document}

\title{Implementation of an adaptive energy-efficient MAC protocol in OMNeT++/MiXiM}

\author{\IEEEauthorblockN{Van-Thiep Nguyen,
Matthieu Gautier,
Olivier Berder}
\IEEEauthorblockA{University Rennes 1, IRISA, INRIA\\
6 Rue Kerampont, 22300 Lannion, France\\
\{van-thiep.nguyen, matthieu.gautier, olivier.berder\}@irisa.fr}}

\maketitle

\begin{abstract}
In recent years, many MAC protocols for wireless sensor networks have been proposed and most of them are evaluated using Matlab simulator and/or network simulators (OMNeT++, NS2, etc). However, most of them have a static behavior and few network simulations are available for adaptive protocols. Specially, in OMNeT++/MiXiM, there are few energy-efficient MAC protocols for WSNs (B-MAC \& L-MAC) and no adaptive ones. To this end, the TAD-MAC (Traffic Aware Dynamic MAC) protocol has been simulated in OMNeT++ with the MiXiM framework and implementation details are given in this paper. The simulation results have been used to evaluate the performance of TAD-MAC through comparisons with B-MAC and L-MAC protocols.
\end{abstract}

\IEEEpeerreviewmaketitle

\section{Introduction}
In Wireless Sensor Networks (WSNs), the MAC layer plays an important role in improving the lifetime of sensor nodes by efficiently managing the communication between these nodes. During the last recent years, many energy-efficient MAC protocols have been proposed to reduce the energy consumption of sensor nodes. One of the important techniques used is duty cycling that refers to change nodes between awake and sleeping states to save energy. B-MAC \cite{Polastre:2004:VLP:1031495.1031508}, WiseMAC \cite{El-Hoiydi:2004:WUL:1126253.1129805} (sender initiated approach) and RI-MAC \cite{Sun:2008:RRA:1460412.1460414}, PW-MAC \cite{Tang11pw-mac:an} (receiver initiated approach) are based on this mechanism.

In asynchronous WSNs, collisions and idle listening are the main sources of energy consumption. In \cite{AlamBMS12}, MM Alam \textit{et al.} proposed an adaptive energy-efficient MAC protocol: Traffic Aware Dynamic (TAD) MAC protocol to reduce the idle listening. In TAD-MAC, the sensor nodes try to dynamically adapt their wakeup-interval  with the amount of received traffic (\textit{i.e.} data packets). In receive node, the data transmission rate of a transmit node is estimated by using a Traffic Status Register (TSR) to update continuously its wakeup-interval. In \cite{AlamBMS12}, the results show that TAD-MAC outperforms other protocols but the simulation have been performed only with Matlab. In this paper, the implementation of TAD-MAC in OMNeT++ with the MiXiM framework is presented. The results show that TAD-MAC can be evaluated easier and more exactly than with Matlab by changing the network topology, adding more sensor nodes and variable traffic flows.

\section{TAD-MAC protocol design}
\label{sec:protocol}
\begin{figure}[!t]
\centering
\includegraphics[width=3.5in]{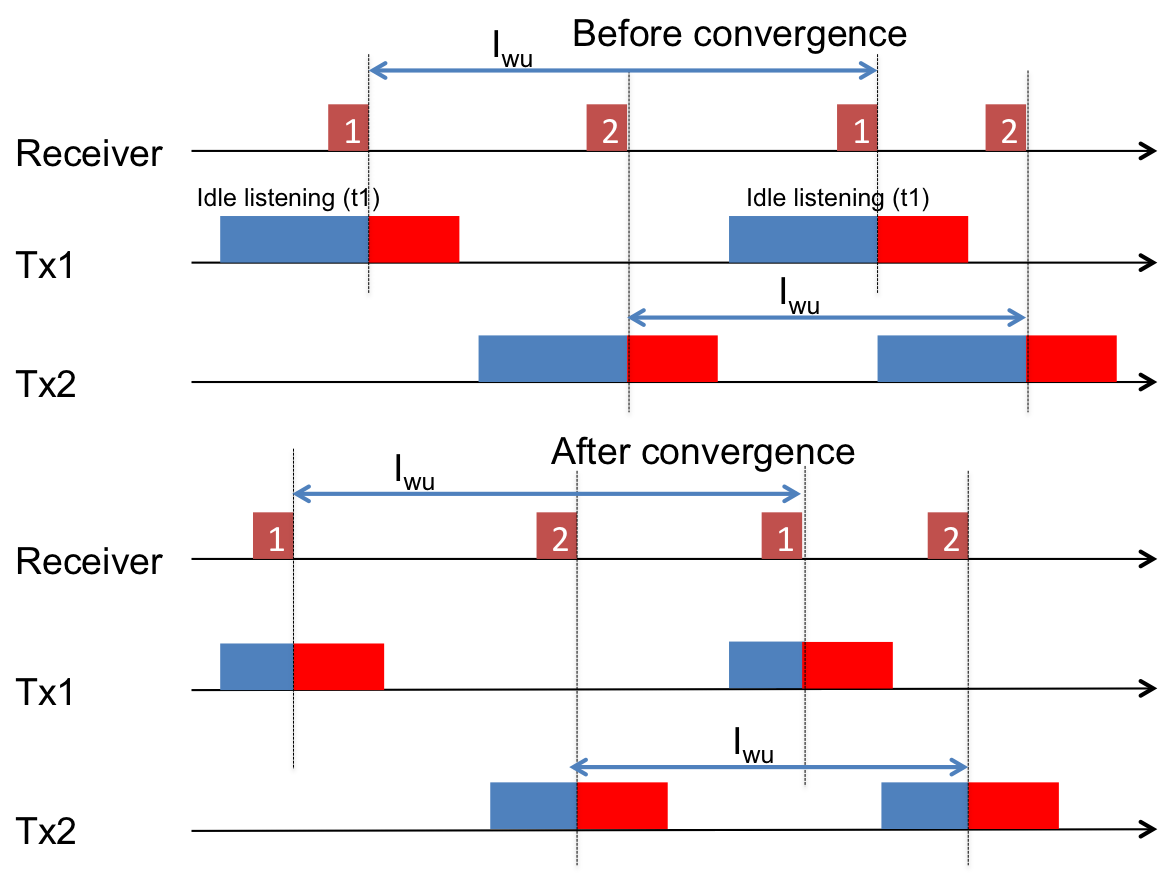}
\caption{The principle of TAD-MAC.}
\label{fig:tadmac}
\end{figure}

The TAD-MAC protocol can be considered in the class of preamble sampling MAC protocol with the receiver initiated strategy. The basic principle of this protocol is described in Fig. \ref{fig:tadmac}. The figure is splitted into two phases (i.e., before convergence and after convergence) that outline the results of a simple network with two nodes (Tx1 and Tx2) attempting to transmit data to a receive node.

During the evaluation phase, before reaching the convergence, the receive node try to adapt its wakeup-interval ($I_{wu}$) to the data transmission rate of each transmit node. Each time the receive node wakes up, it sends the beacon packet to the node which has the nearest wake-up time. The transmit node waits for the beacon signal from the receive node before sending its data. The beacon transmitted contains the specific node identification (node ID). Other intending transmit nodes continue to wait for receiving their beacon.

After several wakeups, the receive node will adapt its $I_{wu}$ based on the statistics of traffic that it receives from each individual transmit node. The second phase, after convergence, indicates that the $I_{wu}$ of receive node has been adapted to traffic of each transmit node in a way that the idle listening is minimal.
 
\begin{figure}[!t]
\centering
\includegraphics[width=3.5in]{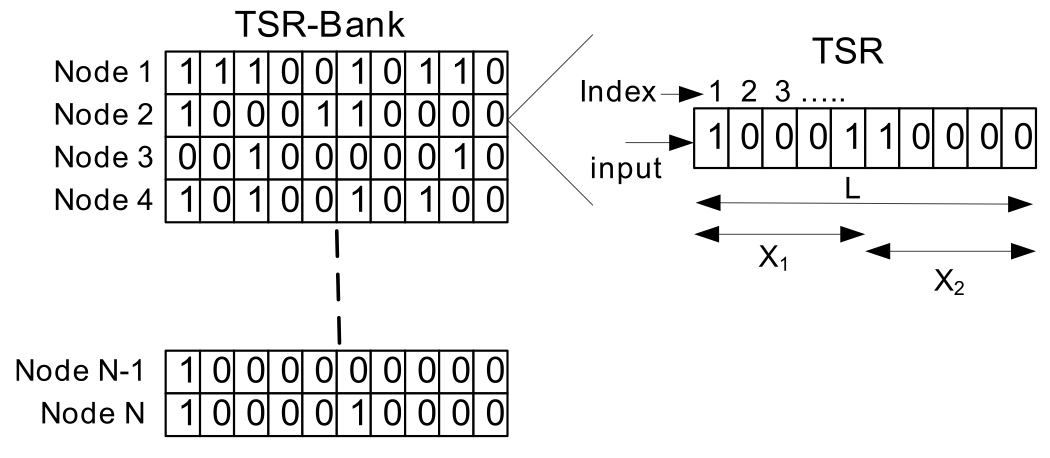}
\caption{TSR-bank: it contains \textit{N} registers for \textit{N} neighbor-nodes.}
\label{fig:tsrbank}
\end{figure}

The TSR is used for each node to collect the status information of the traffic at each node. On the wake-up of receive node, it sends a wake-up beacon to the chosen transmit node, if it receives a data response, the TSR corresponding with this transmit node is filled with a '1', contrarily, the TSR is filled with a '0'. The register contents are shifted one bit right for inserting a new status value in the 1st index. The receive node has a traffic status register bank as shown in Fig. \ref{fig:tsrbank} to keep the traffic status of all the transmit nodes.

After sending the wake-up beacon and updating the TSR, the receive node calculates its next wake-up interval for the time instant ($ t_{i+1} $) by the following adaptive function:
\begin{equation}
I_{wu}(t_{i+1}) = I_{wu}(t_{i}) + [\mu(t_i) + e(t_i)] . t_{ref}
\end{equation}
where \textit{$\mu$} is the output of the weighting average algorithm, \textit{e} is the correlation error and \textit{$t_{ref}$} is defined with reference to the system/simulator clock. The update factor \textit{$\mu(t_i)$} is computed based on two weighted values \textit{$X_1(t_i)$} and \textit{$X_2(t_i)$} shown in Fig. \ref{fig:tsrbank} as
\begin{equation}
\mu(t_i) = \alpha . X_1(t_i) + (1 - \alpha) . X_2(t_i)
\end{equation}
where \textit{$\alpha$} is a constant weighting factor and \textit{$X_1(t_i)$}, \textit{$X_2(t_i)$} are defined as
\begin{equation}
X_1(t_i) = \frac{N_{0,1}}{L/2} . Nc_{0,1} - \frac{N_{1,1}}{L/2} . Nc_{1,1}
\end{equation}
\begin{equation}
X_2(t_i) = \frac{N_{0,2}}{L/2} . Nc_{0,2} - \frac{N_{1,2}}{L/2} . Nc_{1,2}
\end{equation}
where L is the length of the TSR, \textit{$N_{0,1}$}, \textit{$N_{1,1}$}, \textit{$Nc_{0,1}$}, \textit{$Nc_{1,1}$} are the number of zeros, number of ones, number of pairs of zeros, number of pairs of ones in sub-sequence \textit{$X_1$}, whereas \textit{$N_{0,2}$}, \textit{$N_{1,2}$}, \textit{$Nc_{0,2}$}, \textit{$Nc_{1,2}$} are the number of zeros, number of ones, number of pairs of zeros, number of pairs of ones in sub-sequence \textit{$X_2$}.
\section{Implementation in OMNeT++/MiXiM}
\label{sec:implementation}
In this section, the implementation details of TAD-MAC protocol in a network simulator are explained. One of the network simulator which is supporting wireless sensor networks is OMNeT++/MiXiM. OMNeT++ is an object-oriented modular event network simulator and it provides the infrastructure for writing simulations with the component architecture. All elements are called modules and the modules can be connected via gates (or ports) and communicate by exchanging messages. MiXiM is a modeling framework for wireless network (WSNs, body area network, ad-hoc network, vehicular network, etc). There are some wireless MAC protocols that are already implemented in the framework (B-MAC, L-MAC or Zigbee) but no adaptive protocols.

\begin{figure}[t]
\includegraphics[width=3.5in]{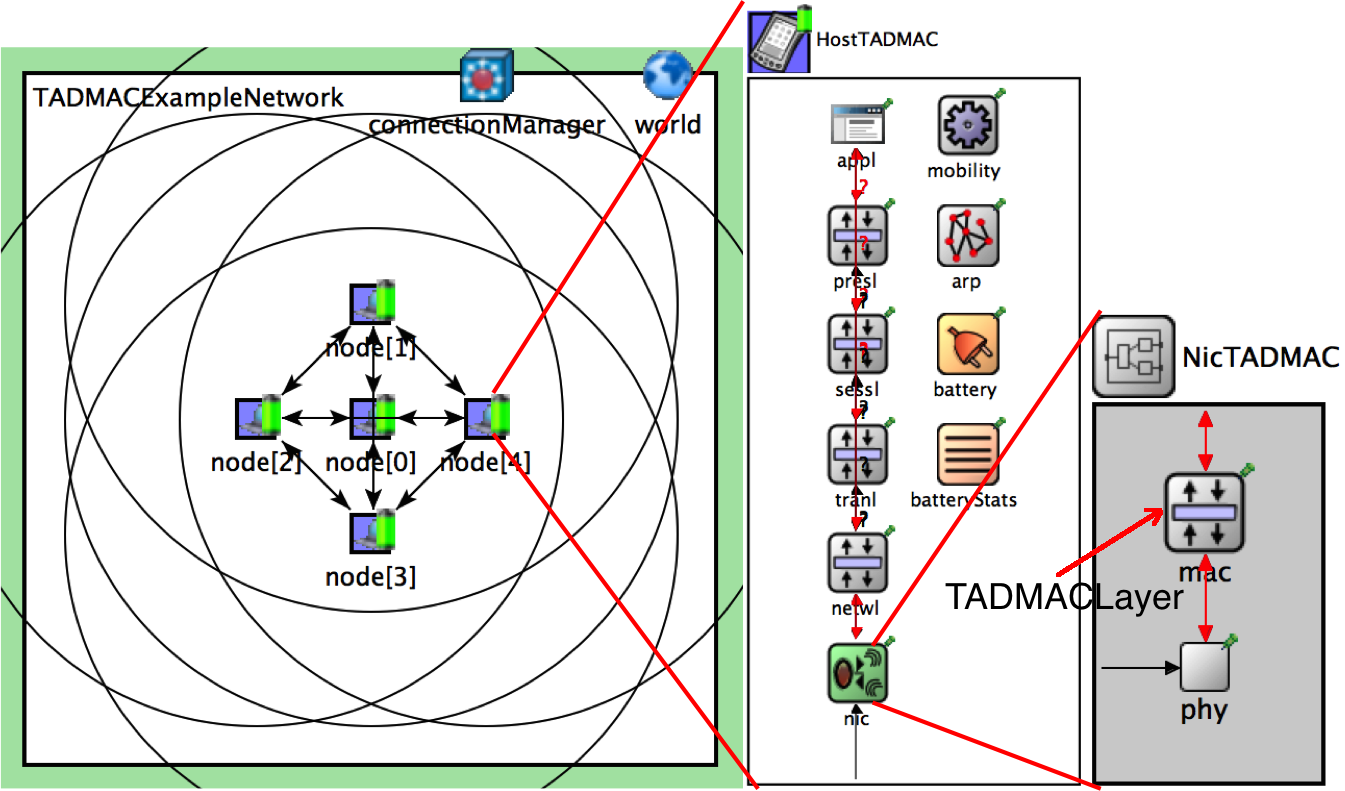}
\caption{A network of \textit{HostTADMAC}.}
\label{fig:host}
\end{figure}

In OMNeT++/MiXiM, there is a definition of network node called \textit{WirelessNodeBattery} which contains the specific elements for a sensor node such as the network interface \textit{WirelessNicBattery}, the power supply \textit{SimpleBattery} and the module \textit{BatteryStats} which calculates the consumption of energy for each element. The main  implementation of TAD-MAC protocol is \textit{\textbf{TADMACLayer}}, which is a module of the MAC layer that is integrated in a new network interface \textit{NicTADMAC} (based on \textit{WirelessNicBattery}). A new host \textit{HostTADMAC} (based on \textit{WirelessNodeBattery}), which contains this network interface is defined as shown in Fig. \ref{fig:host}. Creating a network of \textit{HostTADMAC} and changing the parameters of the network provides are two useful ways to evaluate the performance of TAD-MAC.

\begin{figure*}[!t]
\begin{tabular}{l}
\centerline{
\subfloat[transmit node.]{
  \includegraphics[width =3.5in]{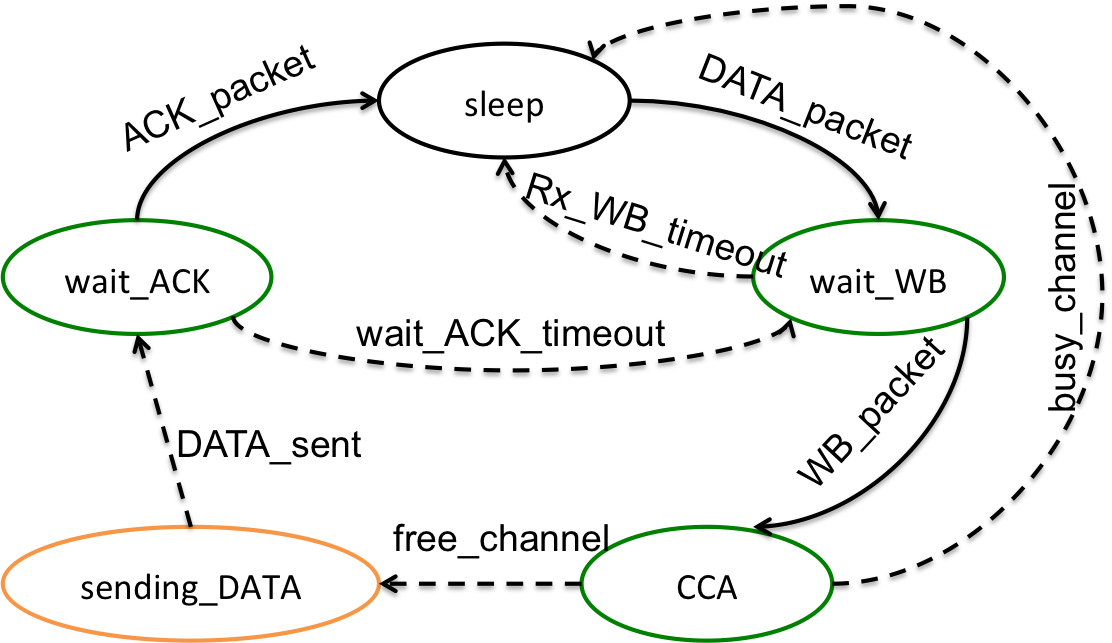}
  \label{fig:sender}
}
\subfloat[receive node.]{
  \includegraphics[width=3.5in]{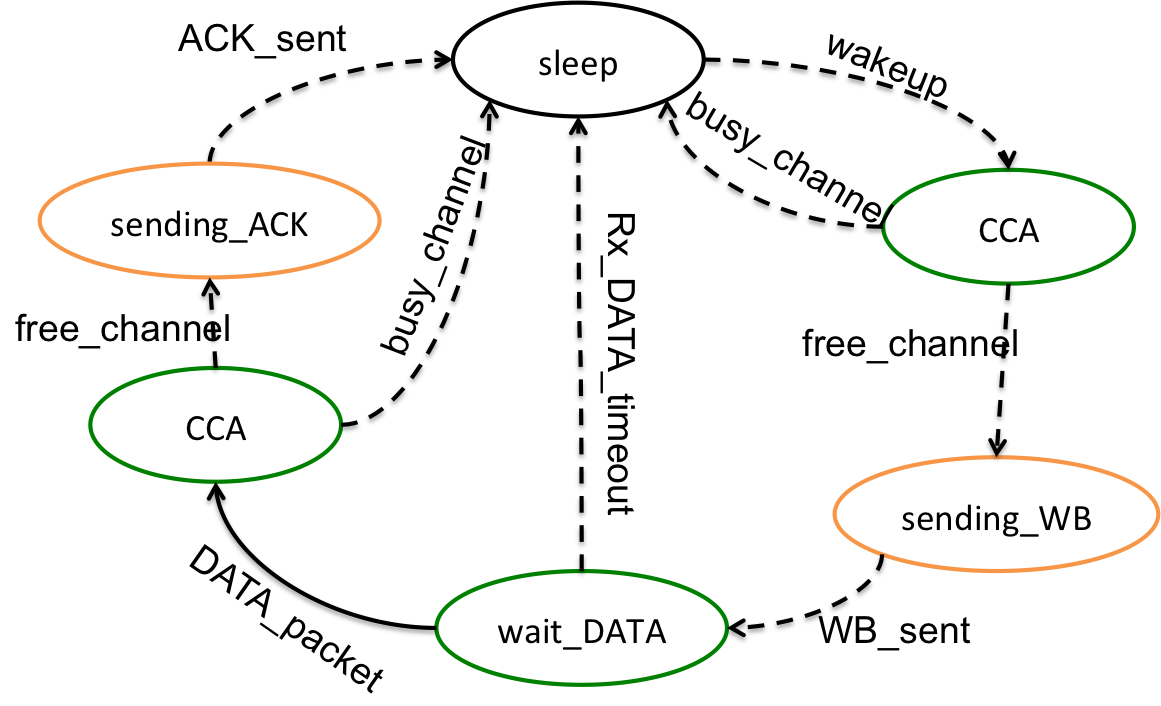}
  \label{fig:receiver}
}}\\
\centerline{\includegraphics[width=3.5in]{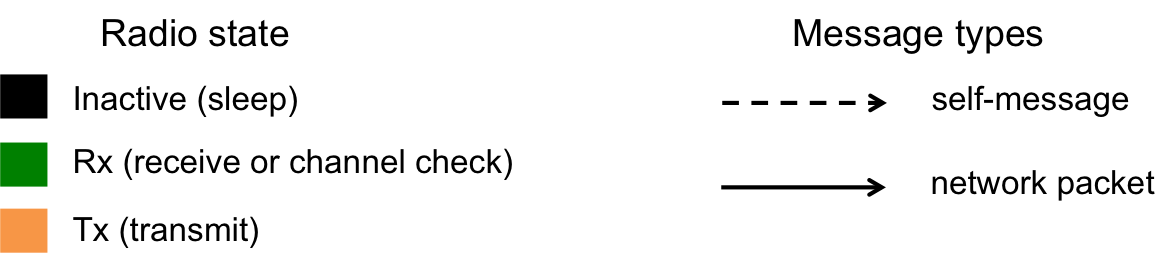}}
\end{tabular}
\caption{State machine of \textit{HostTADMAC}.}
\label{fig:statemachine}
\end{figure*}

As the other modules in MiXiM, \textit{\textbf{TADMACLayer}} contains the gates to connect with other modules/other layers. The messages are used not only as the packets to send between the layers of a sensor node or between two nodes but also as the events to change the state of a node. The state machines of transmit node and receive node shown in Fig. \ref{fig:sender} and Fig. \ref{fig:receiver} describe all the self-messages and the network packets that are used in \textit{\textbf{TADMACLayer}} to change the state of a \textit{HostTADMAC} node. Each time a \textit{HostTADMAC} node wakes up, based on the message that it receives, it will act as transmit node or receive node. A \textit{DATA\_packet}, which is received from network upper layer, signifies that it has data to send so this node will play the role of transmit node. In the contrast, if the message received is the self-message \textit{wakeup}, this node will work as receive node.

In \textit{\textbf{TADMACLayer}}, the \textit{DATA\_packet}, which is sent from network upper layer, is handled in the function \textit{\textbf{handleUpperMsg}}. In this function, after receiving a \textit{DATA\_packet}, a signal is sent to the physicallayer below to switch on the radio antenna if it is in inactive mode or change to \textit{Rx} state. On the other hand, the function \textit{\textbf{handleLowerMsg}} is used to manipulate the data packet, which is sent from physical layer. This packet is also a data packet from the neighbor sensor node such as the \textit{WB\_packet}, \textit{ACK\_packet} for transmit node and the \textit{DATA\_packet} for receive node. The \textit{\textbf{TADMACLayer}} communicates with the physic layer below through another gate to exchange the control message. This type of message, which contains the state information of radio antenna such as the transmission of one packet is over or the radio antenna switching (to \textit{Rx} or \textit{Tx} state) is over, is handled in function \textit{\textbf{handleLowerControl}}. Another important function of \textit{\textbf{TADMACLayer}} is \textit{\textbf{handleSelfMsg}}, which manages the state of MAC layer depending on the type of the received self-message.
\section{Performance evaluation}
\begin{figure*}[t]
\centerline{
\subfloat[results of Matlab simulation.]{
  \includegraphics[width =3.5in]{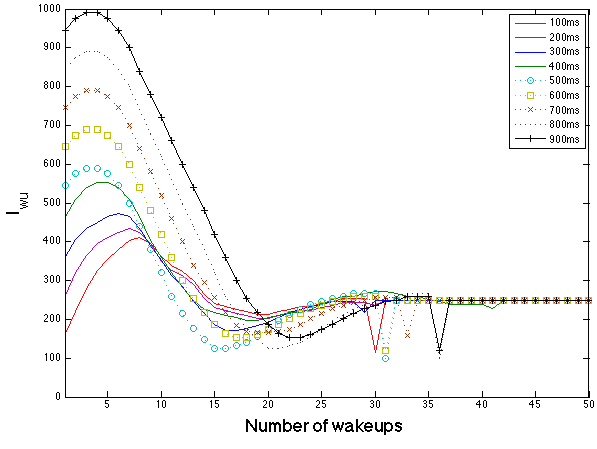}
  \label{fig:result_matlab}
}
\subfloat[results of OMNeT++/MiXiM.]{
  \includegraphics[width =3.5in]{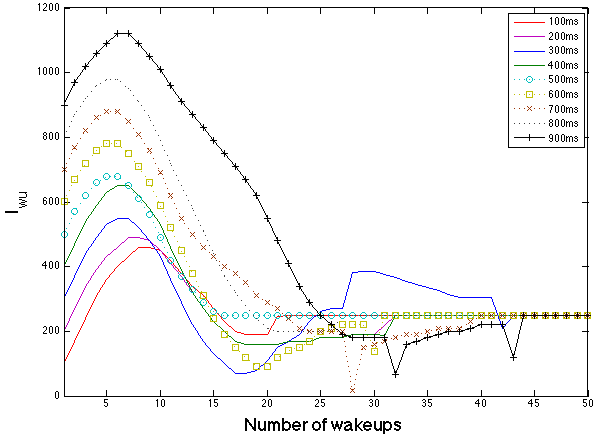}
  \label{fig:result_omnet}
}}
\caption{Behavior of $I_{wu}$ adaptation for number of different initial $I_{wu}$.}
\label{fig:result1}
\end{figure*}

In this section, the performance of TAD-MAC protocol is evaluated by comparing with the simulation of Matlab and other MAC protocols which are implemented in OMNeT++/MiXiM. Firstly, comparing with simulation of Matlab, the results shown in Fig. \ref{fig:result1} describe that the implementation of TAD-MAC protocol in OMNeT++/MiXiM work mainly as in Matlab simulation. But in OMNeT++/MiXiM, some parameters of WSNs are calculated such as the time used to switch radio antenna to Rx or Tx or the time for sending a packet between 2 sensor nodes so the results of OMNeT++/MiXiM simulation are slightly different with the results of Matlab simulation. In Matlab simulation, the number of wakeups of receive node, where the receive node reaches the steady state, is from 33 to 43 but with the same set of initial value of \textit{$I_{wu}$}, in OMNeT++/MiXiM simulation, the number of wakeups is from 16 to 44. Moreover, a complete network is created and managed within a single configuration file \textit{omnet.ini} and a sensor node is inserted easily, the network topology is changed more flexibly without any modification of source code. It seems that in the Matlab simulation, there is no way to calculate the energy consumption of sensor nodes but in the contrast, there is a module in OMNeT++/MiXiM which calculate the energy consumption independently of \textit{\textbf{TADMACLayer}}. By using this module, the consumed energy, the spent time of TAD-MAC is calculated in detail for each state of radio antenna like the transmit state, the receive state and the inactive state which are shown in Table \ref{tab:result}.

\begin{figure}[t]
\centering
\includegraphics[width=3.5in]{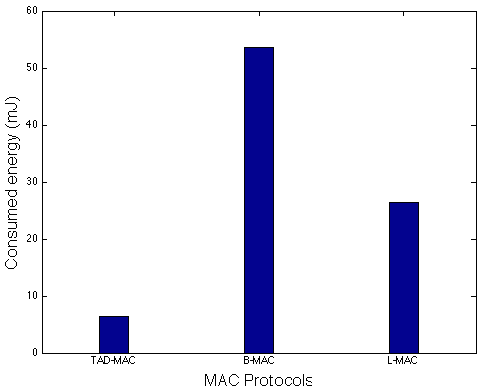}
\caption{Total energy consumption comparison for different MAC protocols (in 1s).}
\label{fig:result}
\end{figure}

Secondly, to evaluate the performance of TAD-MAC, the simulation setup consists of 5 sensor nodes including a coordinating node based on a star topology with 2 types of the data traffic flow: static traffic and variable traffic. The sensor nodes are configured with 3 MAC protocols: B-MAC, L-MAC and TAD-MAC. The evaluation is executed by 100 simulations of 1000s, which are 50 simulations with the static traffic, 50 simulations with variable traffic. The average time spent to transmit, receive and sleep of each node are evaluated in Table \ref{tab:result}. As the results shown in Fig. \ref{fig:result_omnet}, after a number of wakeups, the receive node reaches the steady state. It means that after the convergence, the idle listening is minimized. So during the simulation, the sensor nodes are mostly in inactive state to save energy (nearly 92\% of time), contrarily, the sensor node, which uses B-MAC protocol, spends too much time in sending preamble (about 80\% of time). In L-MAC, the spent time for sending is reduced but the idle listening is still a big problem. The problem of idle listening is resolved completely in TAD-MAC, a sensor node uses about 8\% of time for \textit{Rx} state, which includes also the spent time for clear channel assessment (CCA). The results in Fig. \ref{fig:result} show that the average for energy consumed by TAD-MAC is 5 times less than L-MAC and 10 times less than B-MAC.

\begin{table}[t]
\caption{Energy consumption evaluated for different MAC protocols (in 1s).}
\label{tab:result}
    \begin{tabular}{| l | l | l | l | l |}\cline{1-5}
 & States & B-MAC & L-MAC & TAD-MAC\\ \cline{1-5}
\multicolumn{1}{ |l| }{\multirow{3}{*}{Time (ms)} } &
\multicolumn{1}{ |l| }{Sleep} & 9.6 & 490 & 918 \\ \cline{2-5}
\multicolumn{1}{ |l  }{}                        &
\multicolumn{1}{ |l| }{Rx} & 180 & 497 & 81.03 \\ \cline{2-5}
\multicolumn{1}{ |l  }{}                        &
\multicolumn{1}{ |l| }{Tx} & 800 & 1.9 & 0.53 \\ \cline{1-5}
\multicolumn{1}{ |l| }{\multirow{3}{*}{Energy (mJ)} } &
\multicolumn{1}{ |l| }{Sleep} & 0.0006 & 0.032 & 0.06 \\ \cline{2-5}
\multicolumn{1}{ |l  }{}                        &
\multicolumn{1}{ |l| }{Rx} & 9.78 & 26.89 & 4.38 \\ \cline{2-5}
\multicolumn{1}{ |l  }{}                        &
\multicolumn{1}{ |l| }{Tx} & 45.31 & 0.083 & 0.03 \\ \cline{1-5}
\end{tabular}
\end{table}

\section{Conclusion}
In this paper, the implementation of TAD-MAC in OMNeT++/MiXiM is presented. This is the first adaptive energy-efficient MAC protocol implemented in OMNeT++/MiXiM simulator. The results show that simulation in OMNeT++ has more benefit than Matlab simulation in evaluating the performance of MAC protocol. Via the simulation in OMNeT++, the TAD-MAC protocol is evaluated completely with many test cases of the network topology and variable traffic flows. In all experiments, TAD-MAC significantly outperformed the other MAC protocols. For example, the consumed energy by TAD-MAC is 5 times less than L-MAC and 10 times less than B-MAC in the simulation of 5 nodes with 4 concurrent transmit nodes.

\bibliographystyle{unsrt}
\bibliography{omnetcommunity2014}

\begin{thebibliography}{1}

\bibitem{Polastre:2004:VLP:1031495.1031508}
Joseph Polastre, Jason Hill, and David Culler.
\newblock Versatile low power media access for wireless sensor networks.
\newblock In {\em Proceedings of the 2Nd International Conference on Embedded
  Networked Sensor Systems}, SenSys '04, pages 95--107, New York, NY, USA,
  2004. ACM.

\bibitem{El-Hoiydi:2004:WUL:1126253.1129805}
A.~El-Hoiydi and J.-D. Decotignie.
\newblock Wisemac: An ultra low power mac protocol for the downlink of
  infrastructure wireless sensor networks.
\newblock In {\em Proceedings of the Ninth International Symposium on Computers
  and Communications}, ISCC '04, pages 244--251, Washington, DC, USA, 2004.
  IEEE Computer Society.

\bibitem{Sun:2008:RRA:1460412.1460414}
Yanjun Sun, Omer Gurewitz, and David~B. Johnson.
\newblock Ri-mac: A receiver-initiated asynchronous duty cycle mac protocol for
  dynamic traffic loads in wireless sensor networks.
\newblock In {\em Proceedings of the 6th ACM Conference on Embedded Network
  Sensor Systems}, SenSys '08, pages 1--14, New York, NY, USA, 2008. ACM.

\bibitem{Tang11pw-mac:an}
Lei Tang, Yanjun Sun, Omer Gurewitz, and David~B. Johnson.
\newblock Pw-mac: An energy-efficient predictive-wakeup mac protocol for
  wireless sensor networks.
\newblock In {\em In Proceedings of the 30th IEEE International Conference on
  Computer Communications (INFOCOM 2011}, pages 1305--1313, 2011.

\bibitem{AlamBMS12}
Muhammad~Mahtab Alam, Olivier Berder, Daniel Menard, and Olivier Sentieys.
\newblock Tad-mac: Traffic-aware dynamic mac protocol for wireless body area
  sensor networks.
\newblock {\em IEEE J. Emerg. Sel. Topics Circuits Syst.}, 2(1):109--119, 2012.

\end{thebibliography}

\end{document}